%Paper: hep-th/9207073
%From: holl@dionysos.thphys.ox.ac.uk
%Date: Wed, 22 Jul 92 14:40:13 BST
%Date (revised): Fri, 24 Jul 92 14:44:37 BST

\input panda
%\draftmode{N=2 super Toda.....}
\loadamsmath
\nopagenumbers{\baselineskip=12pt
\line{\hfill OUTP-92-12P}
\line{\hfill hep-th/9207073}
\line{\hfill July, 1992}
\ifdoublepage \bjump\bjump\bjump\bjump\else\vfill\fi
\centerline{\capsone INTEGRABLE N=2 SUPERSYMMETRIC FIELD THEORIES}
\bjump\bjump
\centerline{\scaps Jonathan Evans\footnote{$^1$}
{evansjm\%dionysos.thphys@prg.oxford.ac.uk\newline Supported by SERC
Post-doctoral Research Fellowship B/90/RFH/8856}
and Timothy Hollowood\footnote{$^{2}$}
{holl\%dionysos.thphys@prg.oxford.ac.uk\newline Address after
$1^{\rm st}$ October: Theory Division, CERN, 1211 Geneva 23, Switzerland}}
\sjump
\centerline{\sl Theoretical Physics, 1 Keble Road,}
\centerline{\sl Oxford, OX1 3NP, U.K.}
\vfill
\ifnum\unoduecol=2 \eject\null\vfill\fi
\centerline{\capsone ABSTRACT}
\sjump
\noindent
A classification is given of Toda-like theories with $N{=}2$
supersymmetry which are integrable by virtue of some
underlying Lie superalgebra. In addition to the $N{=}2$ superconformal
theories based on $sl(m,m{-}1)$, which generalize the Liouville model,
a family of massive $N{=}2$ theories based on the algebras
$sl(m,m)^{(1)}$ is found, providing natural generalizations of the
sine-Gordon theory.
A third family of models based on $sl(m,m)$ which have global
supersymmetry, a version of conformal invariance, but no
superconformal invariance is also briefly discussed.
Unlike their $N{=}0$ and $N{=}1$ cousins, the $N{=}2$ massive theories
apparently cannot be directly thought of as integrable deformations of
the corresponding $N{=}2$ superconformal theories. It is shown that
these massive theories admit supersymmetric soliton solutions and a
form for their exact $S$-matrices is conjectured.
\sjump
\ifnum\unoduecol=2 \vfill\fi
\eject
\yespagenumbers\pageno=1
%\doublespaced

\chapter{Introduction: N=1 supersymmetric Toda theories}

\def\m{\mu}

Given a Cartan matrix or a set of simple roots for some Lie algebra, one
can construct an associated integrable bosonic Toda field theory
(see for instance [\Ref{LS}]).
If the algebra is finite-dimensional then the resulting theory is
massless and exhibits an extended conformal symmetry [\Ref{BG},\Ref{HM}]
whilst if the algebra is of
affine Kac-Moody type, then the resulting theory is massive.
In seeking to generalize this construction, it is important to
stress that there is no obvious way to supersymmetrize a given bosonic Toda
theory whilst maintaining integrability.
One can, however, write down integrable Toda theories based on Lie
{\it superalgebras\/} (see [\Ref{O},\Ref{LSS},\Ref{EH}] and references
therein) which contain both bosons and fermions but which
are {\it not\/} supersymmetric in general.
For superalgebras, unlike conventional Lie algebras, there can exist
inequivalent bases of simple roots (up to isomorphism under the Weyl group)
and each of these inequivalent bases leads to a distinct Toda theory.
Each root of a superalgebra carries a ${\Bbb Z}_2$-grading which
makes it either of `bosonic' or `fermionic' type and
it turns out that it is precisely those simple root systems which are
purely fermionic which give rise to supersymmetric Toda theories.

%aims and claims
In this paper we shall consider integrable field theories with $N{=}2$
supersymmetry. We shall extend the treatment of the conformal case
given in [\Ref{EH},\Ref{JE}] and construct
a family of {\it massive\/} $N{=}2$ Toda models which are natural integrable
generalizations of the $N{=}2$ super sine-Gordon theory.
A number of novel features arise compared to the bosonic situation and in
particular these massive theories apparently cannot be interpreted as
integrable deformations of $N{=}2$ superconformal Toda models
in any immediate way.
We will show that the massive $N{=}2$ theories admit supersymmetric
soliton solutions,
which are in fact related to the solitons found recently
[\Ref{tim},\Ref{solsm}] in complex {\it bosonic\/} $sl(m)^{(1)}$ Toda
theories.
In the presence of these solitons the $N{=}2$ supersymmetry algebra acquires
a central term dependent on the topological charge -- in line with
general arguments first given by Witten and Olive [\Ref{olwit}] --
and we describe briefly how this leads to an equation for the
classical soliton masses.
We conclude with some comments and conjectures concerning
$S$-matrices related to these theories.

%start proper

To begin, we summarize in this section the construction of integrable
$N{=}1$ supersymmetric Toda theories. Most of the material is standard,
although we shall treat a number of aspects of the construction
more thoroughly than is usual. This will prove essential in allowing us to
understand some key features of the superalgebra models which have no
counterparts in the bosonic cases.

To write down an integrable theory we require a
{\it contragredient\/} Lie superalgebra (CLSA)
[\Ref{Kac1},\Ref{Kac2},\Ref{Serg}]
which can be defined in terms of a basis of simple roots.
The ${\Bbb Z}_2$-graded commutation relations of the corresponding
step operators can be specified by a Cartan matrix $a_{ij}$ which is
$n {\times} n$-dimensional, symmetric without loss of generality,
and of rank $r$, say.
Specializing immediately to the case of a CLSA with purely fermionic
simple root system, the associated Toda equations are
$$
i D_+ D_- \Phi_i + {1\over\beta} \sum_{j} \m_j a_{ij}\exp \beta \Phi_j
= 0 .
\nfr{eqm}
The $\Phi_j$'s are a set of $n$ scalar superfields
appropriate to two-dimensional super-Minkowski space: they are
functions of the bosonic light-cone coordinates $x^\pm = {1
\over 2} ( t \pm x)$ and of the real fermionic coordinates
$\theta^\pm$, in terms of which each field can be expanded
$$
\Phi_j=\phi_j + i \theta^+ \psi_{j+}  + i \theta^- \psi_{j-} +
i \theta^+ \theta^- F_j .
\efr
The super-derivatives $D_\pm$ are defined by
$$
D_{\pm}= {\del\over\del\theta^{\pm}} -  i \theta^{\pm} \del_{\pm} , \qquad
D^2_{\pm} = - i \del_{\pm},
\efr
and the equations are clearly
invariant under transformations generated by the supercharges
$$
Q_\pm = {\del\over \del \theta^{\pm}} + i \theta^\pm \del_{\pm} , \qquad
Q^2_{\pm} = i \del_{\pm}.
\nfr{qalg}
The coupling constant $\beta$ is dimensionless while the
quantities $\m_i$ are non-zero parameters with the the dimensions of
{\it mass\/} which can be re-scaled by shifting the fields $\Phi_i$.

The integrability of the Toda equations \eqm\ can be established by viewing
them as the zero-curvature conditions for a certain gauge-field in
superspace (for details see [\Ref{O},\Ref{EH}] which follows the
bosonic treatment given in [\Ref{LS}]).
The superfields $\Phi_i$ can take either real or complex values.
It is also consistent, as explained in [\Ref{JE}],
to impose a {\it twisted\/} reality condition of the form
$$
\Phi_i^* = \Phi_{\sigma(i)},
\nfr{twist}
provided that $\sigma$ is a symmetry of order two of the Cartan
matrix, obeying
$$ a_{\sigma(i) \, \sigma(j)} = a_{ij} , \qquad
\m_j^* = \m_{\sigma(j)} , \qquad
\sigma^2 = 1 .
\nfr{ems}
These are the most general possibilities known to be compatible with
integrability and we shall see below that this choice is important in
constructing $N{=}2$ theories.

The number of non-trivial superfield degrees of freedom in the Toda equations
\eqm\ is always equal to $r$, the rank of the Cartan matrix, rather
than to $n$, its dimension.
This follows because if $\xi_j$ are the components of any null
eigenvector of the Cartan matrix, then \eqm\ implies that the field
$ \Phi'=\sum_j\xi_j\Phi_j $ satisfies the free super-wave equation
$D_+ D_- \Phi'=0$.
We may therefore consistently set $\Phi'=0$ and so reduce by one the number
of independent superfields for every independent null eigenvector $\xi$.
It is convenient to regard the remaining Toda fields
as comprising an $r$-dimensional vector $\Phi$. One can then
introduce $n$ constant $r$-dimensional vectors $\alpha_i$ and an
inner-product denoted by a `dot' such that
$$
\Phi_i = \Phi \cdot \alpha_i \, , \qquad
a_{ij} = \alpha_i \cdot \alpha_j \, , \qquad
\sum_i\xi_i\alpha_i=0 \quad \forall\ \xi.
\nfr{const}
The $\alpha_i$'s are actually projections of the simple roots of the
CLSA and the inner-product is similarly induced from the natural
invariant inner-product on the algebra. Although the precise details will not
concern us here, an important point is that this inner-product can
have indefinite signature even when the CLSA is finite-dimensional.
The Toda equations for the independent superfields can now be written
$$
iD_+  D_- \Phi + {1\over\beta} \sum_j
\m_j \alpha_j\exp{\beta\alpha_j\cdot\Phi}
= 0,
\nfr{eqmot}
and they can be derived from the superspace Lagrangian density
$$
L={i\over2} D_+ \Phi \cdot D_- \Phi - {1\over\beta^2}
\sum_{j}\m_j\exp\beta\alpha_j\cdot\Phi.
\nfr{lag}
The character of this theory depends crucially on
the relationship between $r$ and $n$.

If $r = n$, the theory is superconformally invariant. At the classical
level
this symmetry can be realized in superspace by transformations of the
form (see [\Ref{EH}] for details)
$$\Phi \rightarrow \Phi + {2\over\beta}
\rho \log ( D_+ \zeta^+ D_- \zeta^- )
\nfr{scft}
where $\zeta^{\pm}( x^\pm , \theta^\pm)$ and
the vector $\rho$ is defined by $\rho \cdot \alpha_i = \half$
and thus owes its existence to the fact that the $\alpha_j$'s are linearly
independent. A consequence of this symmetry is that $\Phi$ can always
be shifted
so as to set $\mu_j =1$, thereby eliminating any mass parameter from the
theory; we adopt this convention from now on.
A list of all superalgebras of this type was given in
[\Ref{LSS},\Ref{EH}] and we reproduce it here for completeness:
$sl(m,m{-}1) = A(m{-}1,m{-}2)$ $m\geq2$ with $r=2m{-}2$;
$osp(2m{+}1,2m) = B(m,m)$ $m\geq1$ with $r=2m$;
$osp(2m{-}1,2m) = B(m{-}1,m)$ $m\geq1$ with $r=2m{-}1$;
$osp (2m,2m{-}2) = D(m,m{-}1)$ $m\geq2$ with $r=2m{-}1$;
$osp(2m,2m) = D(m,m)$ $m\geq2$ with $r=2m$;
$D(2,1;\alpha)$ $\alpha\in{\Bbb R}\neq0,-1$ with $r=3$.

The alternative is that $r$ is strictly less than $n$.
In this case there is at least one linear relation \const\ obeyed by the
$\alpha_j$'s which is incompatible with the existence of a
superconformal symmetry \scft. Furthermore, the best we can
do by making constant shifts in the superfields $\Phi$ is to fix $r$
ratios of the $\mu_j$'s so that some mass parameter always remains.
A list of the superalgebras of this type
can be found from the work of [\Ref{Serg},\Ref{LSS}]:
$sl(m,m)$ $m\geq2$ with $r=2m{-}2$;
$sl(m,m)^{(1)}$ $m\geq2$ with $r=2m{-}2$;
$sl(2m,2m)^{(2)}$ $m\geq1$ with $r=2m$;
$sl(2m{+}1,2m{+}1)^{(4)}$ $m\geq1$ with $r=2m$;
$sl(2m,2m{-}1)^{(2)} = A(2m{-}1,2m{-}2)^{(2)}$ $m\geq1$ with $r=2m{-}1$;
$osp(2m{+}1,2m)^{(1)} = B(m,m)^{(1)} $ $m\geq1$ with $r= 2m$;
$osp(2m,2m{-}2)^{(1)} = D(m,m{-}1)^{(1)}$ $m\geq2$ with $r = 2m{-}1$;
$osp(2m,2m)^{(2)} = D(m,m)^{(2)}$ $m\geq2$ with $r = 2m{-}1$;
$D(2,1;\alpha)^{(1)}$ with $r = 3$.

Except for the first two families, all entries in the last list
are infinite-dimensional Kac-Moody superalgebras with $n = r{+}1$.
Each is constructed from the given finite-dimensional superalgebra by
means of an outer automorphism of the indicated order.
In all the corresponding Toda models one can shift $\Phi$ so as to
make $\m_j = \m \xi_j$ where $\m$ is a residual mass scale and $\xi_j$
is the {\it unique\/} null eigenvector of the Cartan matrix.
The classical potential of the model then has a minimum at $\Phi = 0$ and
one can see directly that the theory is massive. These cases are
therefore exactly analogous to the bosonic affine theories,
but when we consider instead the first two families on the list
some unfamiliar features emerge.

The superalgebra $sl(m,m)$ is {\it finite\/}-dimensional,
but despite this its Cartan matrix has a single null
eigenvector, implying $n = r{+}1$. The reason
is that this algebra has a one-dimensional centre (which in its defining
representation is generated by the identity matrix).
As we shall see later, the resulting Toda theory has no classical
minimum to its potential so that it is not a massive theory.
Neither is it superconformally invariant, however, because
we have already explained that this relies on the $\alpha_j$'s being
linearly independent. We shall reconcile these apparently conflicting
facts later.
Precisely because the Cartan matrix of $sl(m,m)$ already has one null
eigenvector, the Cartan matrix of its untwisted affine extension
$sl(m,m)^{(1)}$ has two null eigenvectors. This algebra is therefore
unique in having $n=r{+}2$ which will prove important when we search
for $N{=}2$ supersymmetric models. We shall see below that the resulting
Toda theory is massive and has none of the unusual features of the
previous case.

\chapter{N=2 supersymmetric Toda theories}

\def\Qp{ Q^{\prime} }

We now derive conditions for the general $N{=}1$ theory \lag\ to possess
$N{=}2$ supersymmetry, generalizing the treatment of the conformal
case given in [\Ref{EH},\Ref{JE}].
We look for a second supersymmetry
transformation of the field $\Phi$ defined by
$$
\Qp_{\pm} \Phi = J D_{\pm} \Phi,
\nfr{defq}
where $J$ is some matrix which must be compatible with the chosen
reality properties of $\Phi$. This Ansatz ensures that the new
supersymmetry and the original supersymmetry anti-commute.
If $\Qp_\pm$ are to obey the same algebra \qalg\ as $Q_\pm$
we must have
$$ J^2 = -1.
\nfr{iso}
For there to be no
change in the action, which is the superspace
integral of \lag, it
is clear that the kinetic and potential terms must be separately
invariant under this transformation.
The variation of the kinetic terms
is a total superspace derivative if and only if $J$ is antisymmetric with
respect to the inner-product:
$$
\Phi \cdot (J \Phi^{\prime}) = - (J \Phi) \cdot \Phi^{\prime}.
\nfr{asymm}
Given this, the potential terms will also vary into
superspace derivatives if and only if each $\alpha_j$ is
an eigenvector of $J$:
$$
J \alpha_j = \lambda_j \alpha_j.
\nfr{eigen}
The compatibility of $J$ with the reality properties of $\Phi$
now amounts to some set of compatibility conditions for the
eigenvalues $\lambda_j$.

The equations \iso, \asymm\ and \eigen\ are
necessary and sufficient conditions for the
existence of a second supersymmetry in any $N{=}1$ Toda theory,
whether massless or massive.
On combining them we obtain the equivalent conditions
$$
(\lambda_i + \lambda_j) a_{ij} = 0 \, , \qquad \lambda_j^2 = -1 \, .
\nfr{evals}
These are very restrictive:
since $\lambda_j \neq 0$, a second supersymmetry requires
that all the diagonal entries of the Cartan matrix must vanish
and from the lists of $N{=}1$ theories given above one finds that the
only algebras which meet this criterion are as follows.
\hfil \break
(1) $sl (m , m{-}1)$ $(m \geq 2)$ with $n=r=2m{-}2$
and Cartan matrix
$$
a=\pmatrix{0  &  1  &    &    &     &    \cr
           1  &  0  & -1  &    &     &    \cr
             & -1  &  0  &  \ddots  &     &    \cr
              &     & \ddots    &\ddots     & -1     &    \cr
              &     &     & -1    &  0  & 1 \cr
              &     &     &     & 1  &  0 \cr},
\nfr{cmaone}
(2) $sl(m,m)$ $(m\geq2)$ with $n=2m{-}1$. The Cartan matrix is
$$
a=\pmatrix{0  &  1  &     &    &     &        \cr
           1  &  0  & -1  &    &     &        \cr
              & -1  &  0  & \ddots &     &    \cr
              &     &\ddots&\ddots&  1  &     \cr
              &     &     &  1  &  0  &  - 1  \cr
              &     &     &     &  - 1  &  0  \cr},
\nfr{cmatwo}
and it has a unique null eigenvector $(1,0,1, \ldots ,0,1)$ so $r=2m{-}2$.
\hfil \break
(3) $sl(m,m)^{(1)}$ $(m\geq2)$ with $n=2m$. The Cartan matrix is
$$
a=\pmatrix{0  &  -1  &      &       &     &  1 \cr
         - 1  &  0  &  1   &       &     &     \cr
              &  1  &  0   & \ddots&     &     \cr
              &     &\ddots& \ddots&   1 &     \cr
              &     &      &     1 &   0 &  -1 \cr
            1 &     &      &       &  -1 &  0  \cr},
\nfr{cmathree}
and it has two linearly independent
null eigenvectors $(1,0,1,\ldots,1,0)$ and $(0,1,0,\ldots,0,1)$ so
that $r=2m{-}2$.
\hfil \break
In each of these cases there is a unique solution to the
conditions \evals\ given by
$$
\lambda_j = \pm i (-1)^j.
\nfr{lambs}
To complete the analysis, we must now determine to what extent
this solution is compatible with the various reality properties we may wish
to choose for $\Phi$.

If, for any of the above algebras, the superfield $\Phi$ is real,
then the solution \lambs\ is clearly
inconsistent with \defq\ and \eigen\ and so there is no second supersymmetry
in these cases.
It is true that with the choice $\lambda_j = (-1)^j$, which is
consistent with $\Phi$ real, then
\defq\ and \eigen\ would define a new fermionic invariance of
the theory. But this invariance would {\it not\/} be a {\it bona fide\/}
supersymmetry because it would not obey the characteristic algebra \qalg.
(Perhaps such an invariance has interesting consequences, but we shall not
pursue this idea here.)
Another possibility is that we take $\Phi$ to be a complex-valued
superfield, and then \lambs\ is clearly consistent.
Each of the complex Toda theories based on the above algebras is
therefore $N{=}2$ supersymmetric. The last possibility,
which in a number of respects turns out to be the most interesting,
is that $\Phi$ obeys some condition of the form \twist\ for a
non-trivial symmetry $\sigma$. To determine
whether such a choice is possible and whether \lambs\ is then
compatible with it, we will have to
examine the Toda models based on each of these algebras in more detail.

The case of $sl(m,m{-}1)$ has been dealt with previously in [\Ref{JE}]
but we summarize it here for completeness.
There is a non-trivial reality condition
$$
\alpha_j\cdot\Phi^*=\alpha_{\sigma(j)}\cdot\Phi,\qquad\sigma(j)=
2m-1-j , \qquad j = 1 , \ldots , 2m{-}2
\nfr{rcone}
where $\sigma$ is clearly a
symmetry of the Cartan matrix \cmaone\ and \ems\ is satisfied because
we have chosen $\m_j = 1$.
It is easy to see that \lambs\ is compatible with this reality
condition so that there is indeed an additional supersymmetry.
The resulting theory is actually $N{=}2$
superconformally invariant and one can deduce the value of
the (quantum) central charge of the Virasoro algebra
[\Ref{EH}] to be $c=3(m{-}1)\left(1+\frac{m}{\beta^2}\right)$,
where $\beta$ is the coupling constant.
In particular the case $m{=}2$ is the $N{=}2$ super-Liouville theory
and with the choice
$\beta^2= - (k{+}2)$ for $k = 0, 1, 2, \ldots $ one recovers the
$N{=}2$ unitary discrete series of allowed central charges [\Ref{BFK}].
For $m{>}2$ we expect similar ranges of values of the central charge to
correspond to the unitary discrete series for certain $N{=}2$ super
$W$-algebras.

It is instructive to consider the purely bosonic sectors of these
models, which can be calculated by reducing the superspace
Lagrangian to components and eliminating the auxiliary fields.
On general grounds one expects the result to be a
number of decoupled bosonic Toda
theories and this is precisely what appears.
For $m{=}2$, the $N{=}2$ super-Liouville theory,
the bosonic part consists of the usual bosonic Liouville theory
together with an additional free scalar field.
More generally the $sl(m,m{-}1)$ theory leads to a direct sum of
bosonic conformal Toda theories based on
$sl(m)$ and $sl(m{-}1)$ together with one free boson.
The twisted reality conditions
necessary for $N{=}2$ supersymmetry imply similar
twisted reality conditions on the bosonic sub-theories.

Turning next to the algebra $sl(m,m)$, we find that
no non-trivial condition of the form \twist\ is allowed, because
there is no non-trivial symmetry of the Cartan matrix \cmatwo.
For this algebra then, it is only the complex Toda theory which is
$N{=}2$ supersymmetric.
We have alluded previously to the fact that any Toda theory based on
$sl(m,m)$ has some rather bizarre properties; now that we have
written down the Cartan matrix explicitly it is appropriate to
elaborate on these points. Many of them can be traced to the
particular form of the unique null eigenvector $\xi_j$ following
\cmatwo.

First, we see that we have a linear relation
$\sum_{j=1}^m \alpha_{2j-1} = 0$ which is explicitly
incompatible with the existence of a vector $\rho$ obeying $\rho \cdot
\alpha_j = \half$. Hence, as stated earlier, it is
definitely {\it not\/} possible to define superconformal
transformations \scft\ which are symmetries of such a theory.
On the other hand, the theory is {\it not\/} massive, because
the following argument shows that there is no
classical minimum to the potential.
If there were a minimum, we could certainly shift $\Phi$ so that it
occurred at $\Phi = 0$.
But since all the even entries of the null eigenvector
$\xi_j$ vanish, it is impossible to shift the fields so as to make
$\m_j = \mu \xi_j$ for some $\mu$, because such shifts can only
{\it rescale\/} the $\mu_j$.
The linear term in the expansion of the potential about $\Phi =0$ can
therefore never vanish and so there can never be a minimum at this
point.

This puzzling situation can be clarified to some extent by examining
the component content of the simplest example: the theory based on
$sl(2,2)$ with two independent real superfields. On
eliminating auxiliary fields the resulting Lagrangian can be written
in terms of two real
bosons $\phi_1 , \phi_2$ and two real fermions $\psi_{1\pm} , \psi_{2\pm}$
in the form
$$\eqalign{
L & = L_1 - L_2 + i \psi_{2+} \del_- \psi_{1+} - i \psi_{2-} \del_+
\psi_{1-}
\cr
& \quad - i \psi_{1+} \psi_{1-} \exp{\beta \over 2} (\phi_1 + \phi_2)
 - 2 i \psi_{2+} \psi_{2-} \cosh {\beta \over 2} (\phi_1 - \phi_2) .
\cr }\nfr{weird}
Here $L_j$ denotes the Liouville Lagrangian for the bosonic field
$\phi_j$ with coupling constant $\beta$ so that the bosonic sector
of the model is conformally invariant in the standard way.
By inspection, this conformal symmetry can be uniquely extended to the
whole action, but only if the fermions transform with non-standard
{\it integer\/} conformal weights:
$$
\psi_{1\pm} \rightarrow (0,0) , \qquad
\psi_{2+} \rightarrow (1,0) , \qquad \psi_{2-} \rightarrow (0,1) .
\efr
Such an assignment clearly implies that the supercurrent for
the $N{=}1$ {\it global\/} supersymmetry of the model cannot have the
conformal dimensions required to extend this non-standard conformal
symmetry to a superconformal symmetry.
Thus we are able to reconcile
the co-existence of a global supersymmetry and a version of
conformal invariance with the absence of superconformal invariance.
This is also consistent with the fact that one
can eliminate all mass parameters from the theory, as we have done in
\weird, by redefinitions of the various {\it component fields\/},
whereas this cannot be achieved by constant shifts in the original
{\it superfields}, as we remarked earlier.

Although we have considered just the simplest case, similar remarks
should also apply to more complicated models. Thus for the complex Toda theory
based on $sl(m,m)$ we expect that the
supercurrents for the $N{=}2$ {\it global\/} supersymmetry will have exotic
conformal weights with respect to a similar non-standard conformal
invariance. The precise r\^ole played by this non-standard conformal
symmetry is rather obscure at present.
Nevertheless, there are some intriguing features associated with it,
among which is the fact that the spin assignments for the
fermions are those of a ghost system such as might be
encountered in a topological theory.
Perhaps these unusual features would repay a more thorough study.

Finally we come to a detailed examination of the
Toda theories based on $sl(m,m)^{(1)}$. Starting from the general action \lag,
the existence of two null eigenvectors of the Cartan matrix \cmathree\
means that there are two independent mass parameters in the
theory which cannot be absorbed away by shifting $\Phi$.
More precisely, the particular form of the null eigenvectors means
that we can set $ \m_{2j-1} = \m $ and $ \m_{2j} = \m'$, say.
The Lagrangian then reads
$$
L = {i\over2} D_+ \Phi \cdot D_- \Phi
- {\m\over\beta^2} \sum_{j=0}^{m{-}1} \exp \beta \alpha_{2j{+}1} \cdot \Phi
\, -{\m'\over\beta^2} \sum_{j=0}^{m{-}1} \exp \beta
\alpha_{2j} \cdot \Phi,
\nfr{lagaff}
and it is clear that the classical minimum occurs at $\Phi = 0$.
The resulting theory is massive, and it displays
none of the peculiarities exhibited by the algebraically related
model based on $sl(m,m)$.

Despite the occurrence of {\it two\/} parameters $\m$ and $\m^\prime$
with the dimensions of mass, the classical masses of the
elementary excitations of the theory
depend just on the product $\m \m^{\prime}$ and the ratio
$\m/\m^\prime$ appears only in the higher order couplings between bosons
and fermions. It is therefore more accurate to describe this model as
having a single mass scale together with an extra dimensionless
coupling constant $\mu / \mu^\prime$ in addition to $\beta$.
We should stress also that the masses mentioned above are always real and
positive even though the kinetic energy contains terms of negative
sign in general (because of the indefinite nature of the superalgebra
inner-product).

We must now consider the allowed reality choices for $\Phi$ in this
case. There is certainly a non-trivial symmetry of \cmathree\
which we can use to write down a reality condition
$$
\alpha_j \cdot \Phi^* = \alpha_{\sigma(j)} \cdot \Phi , \qquad \sigma
(j) = 2m - 1 - j , \qquad j = 0, \ldots , 2m{-}1
\nfr{rctwo}
(Note the range of labelling chosen here and in \lagaff\ which will
turn out to be convenient later.)
But this is consistent with the Toda equations
only if we set $\m^{\prime}{=}\m$ so as to fulfill \ems, thereby
fixing the value of the additional dimensionless parameter discussed
above.
Having done so, it is easy to see that \rctwo\ is also compatible with
\lambs\ so that we have found another class of $N{=}2$ supersymmetric
theories.

For $m{=}2$ the above construction gives the
$N{=}2$ super sine-Gordon model which contains as its bosonic limit
decoupled copies of both the sine- and sinh-Gordon theories.
The fact that this really is the natural generalization of the
conventional bosonic sine-Gordon theory is best appreciated in the
language of $N{=}2$ superspace which we shall introduce in the next
section. Some previous work linking integrability of this model with the
algebra $sl(2,2)^{(1)}$ can be found in [\Ref{KI}] and a recent discussion
of its $S$-matrix is given in [\Ref{ssg}].

For $m{>}2$ we have found the natural multi-component integrable
generalizations of the N{=}2 sine-Gordon theory.
In the general case the bosonic limit consists of a direct sum of two
bosonic Toda theories, each based on $sl(m)^{(1)}$.
The twisted reality conditions on the superfields
descend to these bosonic sub-theories in a way which is not entirely
trivial. Nevertheless, the mass spectrum of these bosonic theories is
independent of the reality condition used [\Ref{JE}] and this fact provides a
simple way of seeing that the mass spectrum of the superalgebra model
is real and positive despite the possibility of indefinite kinetic
energy. Details of how the reality conditions descend will be given
elsewhere.

\chapter{N=2 Superspace}
\def\thetab{\bar \theta}
\def\Db{\overline D}

We shall now take the extended supersymmetric theories
constructed from $sl(m,m{-}1)$ and $sl(m,m)^{(1)}$ by using
the reality conditions \rcone\ and \rctwo\ and re-formulate them in
$N{=}2$ superspace.
In standard fashion, we can introduce an $N{=}2$ superspace by taking
{\it complex\/} fermionic coordinates $\theta^\pm$ and their
conjugates $\thetab^\pm$; the $N{=}1$ superspace we have been
considering thus far is contained as the subspace with fermionic
coordinates $\theta^\pm = \thetab^\pm$.
The $N{=}2$ superspace derivatives are
$$
D_\pm = {\del \over \del \theta^\pm} - {i\over2} \thetab^\pm \del_\pm ,
\qquad
\Db_\pm = {\del \over \del \thetab^\pm} - {i\over2} \theta^\pm \del_\pm
\qquad \Rightarrow
\qquad \{ D_\pm , \Db_\pm \} = - i \del_\pm ,
\efr
with all other brackets vanishing.
A chiral $N{=}2$ superfield $\Psi$
obeys $\Db_\pm \Psi = 0$ and its complex conjugate $\Psi^*$ is
antichiral obeying $D_\pm \Psi^* = 0$.
The generic form of an $N{=}2$ superspace action is
$$
S = \int d^2 x \, d^2 \theta \, d^2 \thetab  \, K ( \Psi^*_j , \Psi_j )
- \int d^2 x \, d^2 \theta \, W ( \Psi_j )
- \int d^2 x \, d^2 \thetab \, W^* ( \Psi^*_j ),
\efr
where $K$ is a hermitian form and $W$ is a holomorphic superpotential.
Notice in particular how the
interaction terms occur in complex conjugate pairs to make the action
as a whole real.

The theories of interest to us have been defined in terms of $2(m{-}1)$
independent, complex $N{=}1$ superfields obeying \rcone\ and \rctwo.
But there is an obvious way to identify a complex $N{=}1$ superfield
with a chiral $N{=}2$ superfield, component by component.
In the cases at hand it
is convenient to introduce $m{-}1$ chiral $N{=}2$ superfields
$\Psi_j$ by means of the identifications
$$
\Psi_j \leftrightarrow \alpha_j \cdot \Phi \quad
j \, \, {\rm odd}, \qquad
\Psi_j \leftrightarrow \alpha_j \cdot \Phi^* \quad
j \, \, {\rm even} , \qquad
j = 1 , \ldots , m{-}1
\efr
which is consistent with \rcone\ and \rctwo.
In both the conformal and massive cases the kinetic terms can now be
written in $N{=}2$ language by choosing
$$
K = \sum_{i, \, j=1}^{m-1} \Psi_i^* \, k^{-1}_{ij} \, \Psi_j ,
\qquad
k = \pmatrix{
0 & 1 &  &      & & \cr
1 & 0 &-1&      & & \cr
  &-1 & 0&      & & \cr
  &   &  &\ddots&   & \cr
  &   &  &      & 0 & (-1)^{m-1} \cr
  &   &  &      & (-1)^{m-1} & (-1)^{m}\cr
}.
\efr
For the conformal theory based on $sl(m,m{-}1)$ the appropriate
superpotential is
$$
W = {1\over\beta^2} \sum_{j=1}^{m-1}  \exp \beta \Psi_j,
\nfr{spone}
while for the massive theory based on $sl(m,m)^{(1)}$ it is
$$
W ={\m\over\beta^2}\left \{
\sum_{j=1}^{m-1}\exp\,\beta\Psi_j
+ \exp \left( - \beta \sum_{j=1}^{m-1} \Psi_j \right) \right \}.
\nfr{sptwo}
Notice that the rank $r$ is even for both algebras,
corresponding to the hermitian structure of the kinetic terms.
It is important for the affine theory that $n$ is also even,
because the general form of an $N{=}2$ superspace action requires that the
exponential interaction terms occur in complex conjugate pairs.
The unique property of the family
$sl(m,m)^{(1)}$ in having $n = r{+}2$ is therefore intimately related
to the extended supersymmetry present in this Toda theory.

The extended superspace formalism makes transparent why the models
based on $sl(2,1)$ and $sl(2,2)^{(1)}$
are the natural $N{=}2$ extensions of the Liouville and
sinh-Gordon theories. They can be formulated using a single chiral
$N{=}2$ superfield and superpotentials
$$
W = {1\over\beta^2}\exp \beta \Psi
\qquad
{\rm and}
\qquad
W = {2\mu\over\beta^2}\cosh\beta \Psi,
$$
respectively, which are obvious generalizations of the bosonic cases.
Due to the algebraic complexities hidden in the superspace notation, however,
the bosonic sectors of such theories are not immediately obvious.
As we stated earlier, the $N{=}2$ Liouville theory reduces to the
bosonic Liouville theory plus one free scalar, whilst the $N{=}2$
sine-Gordon reduces to decoupled copies of the sinh-Gordon and sine-Gordon
theories, on setting the fermions to zero.
We should also point out the $N{=}2$ theories corresponding to higher
values of $m$ are not simple generalizations of bosonic Toda theories.

\chapter{Integrable Deformations and N=2 Toda Theories}

Toda theories provide a framework for discussing integrable
deformations of certain conformal field theories by particular primary
operators in the following way [\Ref{HM},\Ref{sgpert}].
Let $L$ be the Lagrangian for a conformal Toda theory based on some
CLSA $g$ with some choice of simple roots.
Then the massive theory defined by
$$
L + \lambda\cases{\exp(\beta\chi\cdot\Phi)&$
\chi\ {\rm fermionic}$\cr
i \theta^+\theta^-\exp(\beta\chi\cdot\Phi)\quad&$\chi\ {\rm
bosonic},$\cr}
\efr
will still be integrable if the vector $\chi$ appearing in the
perturbing operator extends the simple root system for $g$ to a simple
root system for some affine CLSA.
In the simplest case $\chi$ is the negative of the
highest root for $g$ which corresponds to the additional root of the untwisted
affine CLSA $g^{(1)}$, but more generally the larger root
system may be that of some twisted affine CLSA.
Notice that the deformation preserves or breaks $N{=}1$ supersymmetry
according to whether $\chi$ is graded fermionic or bosonic respectively.

It is natural to ask whether one can find Toda theories describing
integrable perturbations which preserve $N{=}2$ supersymmetry.
For the $N{=}2$ superconformal Toda model based on $sl(m,m{-}1)$, the
simplest deformation corresponding to $sl(m,m{-}1)^{(1)}$
was considered in [\Ref{EH}].
Starting from the basis of purely fermionic simple roots, it was found that
$\chi$ is always bosonic so that the deformation breaks even the
$N{=}1$ supersymmetry of the original theory.
(This analysis assumed standard reality conditions for $\Phi$ but it
carries over immediately to the case of twisted
reality conditions relevant to the genuine $N{=}2$ theories considered here.)
Now that we have analyzed the construction of $N{=}2$ massive theories
more thoroughly, however, we can address the issue of $N{=}2$
deformations more systematically.

Our work above immediately suggests that we might try to interpret
the massive $N{=}2$ theories based on $sl(m,m)^{(1)}$
as $N{=}2$ supersymmetric integrable deformations of the
conformal theories based on $sl(m,m{-}1)$.
These models have the same numbers of fields
(because the ranks of the algebras are equal for fixed $m$) and
the latter theories can be `embedded' in the former in a
natural way, as is evident from the Cartan matrices \cmaone\ and \cmathree.
We are thus led to consider a deforming operator in the $sl(m,m{-}1)$ model
consisting of a sum of {\it two\/} exponentials
$$
\lambda ( \, \exp \beta \alpha_0 \cdot \Phi \,  + \,
\exp \beta \alpha_{2m{-}1} \cdot \Phi \, ),
\quad \,
\alpha_{0} = - \sum_{j=1}^{m-1} \alpha_{2j} ,
\quad \,
\alpha_{2m-1} = - \sum_{j=1}^{m-1} \alpha_{2j{-}1} .
\efr
(the ratio of the coefficients is fixed by the condition of
$N{=}2$ supersymmetry which required $\mu = \mu^\prime$ in \lagaff.)
Unfortunately, a calculation reveals that
the holomorphic and antiholomorphic dimensions of this perturbing
operator are $\frac12(1-m)$, which is always {\it negative\/} (as was
noted for the Sine-Gordon case in [\Ref{ssg}]).
This strongly suggests that such
a simple interpretation purely at the level of Lagrangians
is too na\"\i ve, and so we shall not advocate it here. This is to be
contrasted
with the situation in the Landau-Ginzburg formalism where deformations
can be considered directly at the level of the Lagrangian [\Ref{ntsm}].
A more satisfactory route to clarifying the relationships between
the $N{=}2$ Toda models considered here
would be to investigate the ultra-violet limit of the
quantum $S$-matrix of the affine theories (for recent relevant work
see [\Ref{fend}]).

\chapter{Solitons in massive N=2 Toda theories}

Both the complex and twisted \rctwo\ families of massive
$N{=}2$ theories based on $sl(m,m)^{(1)}$ admit supersymmetric
soliton solutions. Recent
work [\Ref{tim},\Ref{solsm}] on solitons in {\it bosonic\/} Toda
models, however,
suggests that the complex theories may be the most natural
setting in which to study such solutions, so we will concentrate on
these here. In the bosonic case,
solitons of the complex $sl(m)^{(1)}$ Toda equations
have been constructed in [\Ref{tim}]. (These theories are to be
thought of as natural generalizations of the
sine-Gordon model, whereas the conventional real Toda theories
generalize the sinh-Gordon model.)
It turns out that the classical masses of these solitons are real, in
spite of the fact that the Hamiltonian is not hermitian. A form for
the soliton $S$-matrix has also been proposed. This is is generically
non-unitary, as expected, although it seems that for some
particular region for the coupling constant $\beta$, unitarity is
restored [\Ref{solsm}].

Consider the complex $N{=}2$ theory based on the CLSA $sl(m,m)^{(1)}$
and define for $j=1,2,\ldots,m$
$$
\alpha_{2j-1}\cdot\Phi=\Phi_j^{(1)}-i\Phi_j^{(2)},\qquad
\alpha_{2j-2}\cdot\Phi=i\Phi_{j-1}^{(2)}-\Phi_{j}^{(1)},
\efr
with $\Phi_j^{(1,2)}\equiv\Phi_{j+m}^{(1,2)}$.
In looking for classical solutions to such a field theory it is customary
to set all fermions to zero and to consider the resulting purely bosonic
equations. With the definitions above, this implies that
$\phi^{(1)}$ satisfies a bosonic
$sl(m)^{(1)}$ Toda equation with coupling constant $\beta$,
whereas $\phi^{(2)}$ satisfies a complex
$sl(m)^{(1)}$ Toda equation with coupling constant $i \beta$:
$$
\del_+ \del_- \phi^{(2)}_j+{\m^2\over i\beta}
\left(e^{i\beta(\phi^{(2)}_j-\phi^{(2)}_{j+1})}-e^{i\beta(\phi^{(2)}_{j-1}
-\phi^{(2)}_j)}\right)=0.
\efr
We can now use the results of $[\Ref{tim}]$ directly to construct
multi-soliton solutions. For example, the one soliton solutions are
$$
\phi^{(2)}_j(x,t)=-{1\over i\beta}\log\left({1+e^{\rho(x-vt)+{2\pi
ia\over m}j+\xi}\over 1+e^{\rho(x-vt)+{2\pi ia\over
m}(j-1)+\xi}}\right),\qquad \phi^{(1)}_j=0.
\nfr{onesol}
where $\rho$, $\xi$ and $v$ are constants satisfying
$\rho^2(1-v^2)=4\mu^2\sin^2(\pi a/m)$
and $a=1,2,\ldots,m{-}1$. The topological charge of the soliton
\onesol\ is a weight of the $a^{\rm th}$ fundamental representation of
$sl(m)$ and the classical mass of the soliton is equal to
$$
M_a={4\mu m\over\beta^2}\sin\left({\pi a\over m}\right).
\nfr{solmass}

We now show that these one soliton solutions are
`supersymmetric' in the sense that they
are invariant under a particular supersymmetry transformation.
Working in $N{=}1$ superspace, with the second
supercharge being $Q_\pm'=JD_\pm$, as before, we consider the
variation of the soliton solution under the transformation generated by
$\varepsilon^+Q_++\varepsilon^-Q_-+\varepsilon^{+\prime}Q'_++
\varepsilon^{-\prime}Q'_-$.
The change in a general bosonic field configuration $\phi$
is automatically zero because the fermion
fields vanish, so the only non-trivial equations result from requiring
that the variation of the fermions should also vanish. This gives gives
$$\eqalign{
\delta\psi_+&=-\varepsilon^+\partial_+\phi+\varepsilon^{+\prime}J\partial_+\phi
+\varepsilon^-F+\varepsilon^{-\prime}JF\cr
\delta\psi_-&=-\varepsilon^-\partial_-\phi+\varepsilon^{-\prime}\partial_-\phi
-\varepsilon^+F-\varepsilon^{+\prime}JF,\cr}
\efr
where the auxiliary field takes its on-shell value:
$$
F=-{\m\over\beta}\sum_j\alpha_j\exp\beta\alpha_j\cdot\phi .
\efr
One finds that these variations vanish for the restricted transformation
$\varepsilon^-=\lambda\varepsilon^+$,
$\varepsilon^{+\prime}=\sigma\varepsilon^+$ and
$\varepsilon^{-\prime}=\lambda\sigma\varepsilon^+$,
on taking $\phi$ as given in \onesol, because it then
satisfies the first order equation
$$
-(1-\sigma J)\partial_+\phi+\lambda(1+\sigma J)F=0,
\efr
with
$$
\lambda=\sqrt{1-v\over1+v},\qquad \sigma=\tan\left({
\pi a\over2m}+{\pi\over 4}\right).
\nfr{params}

There exists a formula relating the mass
of such a supersymmetric soliton to its topological charge
which can be derived from a careful
treatment of the non-trivial central terms in the superalgebra induced
by such topological configurations [\Ref{olwit},\Ref{sssm}]. In
the present context one finds that this formula for the mass is
$$
M={\mu\over\beta^2}\int dx\,{\partial\over\partial x}
\left({1+i\sigma\over1-i\sigma}
\sum_{j \, \, {\rm even}}e^{\beta\alpha_j\cdot\phi}+
{1-i\sigma\over1+i\sigma}
\sum_{j \, \, {\rm odd}}e^{\beta\alpha_j\cdot\phi}\right),
\nfr{masstop}
where $\sigma$ is the parameter appearing in the supersymmetry
transformation \params. It is straightforward
to verify explicitly that with $\phi$ given by \onesol\ the masses
\solmass\ are correctly reproduced by \masstop. We therefore obtain a
new way of seeing that these soliton masses must be real,
which follows because the two terms in \masstop\ are
complex conjugates. We intend to present the above
calculations in greater detail elsewhere, together with a more complete
discussion of
some related issues.

Although it is is somewhat premature to speculate on the possibilities
for defining a consistent $N{=}2$ massive quantum theory associated to
these classical equations, experience with the bosonic theories is
encouraging in that an exact $S$-matrix for the soliton
solutions has been postulated [\Ref{solsm}].
In the present case with $m{=}2$, giving the $N{=}2$ super
sine-Gordon theory, the $S$-matrix is thought to be
related to the $S$-matrix of the bosonic sine-Gordon theory in a
very simple way [\Ref{ssg},\Ref{fend}]
$$
S^{N=2}_{SG}\left(\beta_{N=2};\theta\right)=S^{N=0}_{SG}\left(\beta_{N=0};
\theta\right)\otimes S^{N=2}[m{=}2](\theta),
\efr
where $S_{SG}$ is the $S$-matrix of the sine-Gordon theories of the
indicated supersymmetry, and $S^{N=2}[m](\theta)$ is one of the
`minimal' $N{=}2$ supersymmetric $S$-matrices constructed in
[\Ref{fend}]. The coupling
constant $\beta_{N=2}$ turns out to be the bare coupling of the bosonic
sine-Gordon theory. The fact that
the final form is a product of a bosonic $S$-matrix and a fermionic
$S$-matrix seems to be a characteristic feature of these types of theory
[\Ref{sssm}]. It seems plausible that the $m>2$
theories would yield a similar structure with the bosonic $S$-matrix
factor being replaced by the $sl(m)$ soliton $S$-matrix of
[\Ref{solsm}] (but without
the factor $S_{\rm min}$, since in this case $S^{N=2}[m]$ would provide the
necessary pole structure). It would clearly be interesting to
investigate the spectrum of states that follows from such an Ansatz,
and also to consider the question of unitarity along the lines of
[\Ref{solsm}]. A proposal for the theory giving this $S$-matrix is
made in [\Ref{fend}], however, it is not the same as our proposal, and its
integrability is questionable given the cautionary remarks we made in
the introduction regarding the construction of {\it integrable\/}
supersymmetric theories.

TJH would like to thank Merton College, Oxford, for a Junior Research
Fellowship. JME would like to thank the U.K. Science and Engineering
Research Council for financial support in the form a Post-doctoral
Research Fellowship.

Note added: after submitting this paper for publication we became
aware of ref.~[\Ref{GPZ}] (we thank M. Grisaru for bringing it to our
attention). These authors consider field theories based on the
algebras $sl(m,m)^{(1)}$ and compute one-loop corrections to particle
masses, although they do not discuss $N{=}2$ supersymmetry. They also
consider theories related to $sl(m,m)$ and claim to find a quantum
superconformal symmetry, at variance with our results. Clearly
these points deserve more detailed investigations. Some additional
references on super Toda theories can be found in
[\Ref{stod1},\Ref{stod2},\Ref{stod3}].

\references

\beginref
\Rref{olwit}{E. Witten and D. Olive, Phys. Lett. {\bf B78} (1978) 97}
\Rref{tim}{T.J. Hollowood, `{\sl Solitons in affine Toda field theories\/}',
Oxford University preprint OUTP-92-04P, Nucl. Phys. {\bf B} {\it to appear\/}}
\Rref{JE}{J.M. Evans, `{\sl Complex Toda theories and twisted reality
conditions\/}', Oxford University preprint OUTP-91-39P, Nucl. Phys.
{\bf B} {\it to appear\/}}
\Rref{solsm}{T.J. Hollowood, `{\sl Quantizing $SL(N)$ solitons and the
Hecke algebra\/}', Oxford University preprint OUTP-92-03P, Int. J.
Mod. Phys. {\bf A} {\it to appear\/}}
\Rref{HM}{T. Eguchi and S-K. Yang, Phys. Lett. {\bf B224} (1989) 373\newline
T.J. Hollowood and P. Mansfield, Phys. Lett. {\bf B226} (1989) 73}
\Rref{sgpert}{A. LeClair, Phys. Lett. {\bf B230} (1989) 103\newline
D. Bernard and A. LeClair, Phys. Lett. {\bf B247} (1990) 309\newline
F.A. Smirnov, Nucl. Phys. {\bf B337} (1990) 156;
Int. J. Mod. Phys. {\bf A4} (1989) 4213\newline
T. Eguchi and S-K. Yang, Phys. Lett. {\bf B235} (1990) 282}
\Rref{ssg}{K. Kobayashi and T. Uematsu, Phys. Lett. {\bf B275} (1992) 361}
\Rref{Kac1}{V.G. Kac, Adv. Math. {\bf26} (1978) 8}
\Rref{Kac2}{V.G. Kac, Adv. Math. {\bf30} (1978) 85;
Springer Lecture Notes in Physics vol.~94 (1979) 441}
\Rref{O}{M.A. Olshanetsky, Commun. Math. Phys. {\bf88} (1983) 63}
\Rref{EH}{J.M. Evans and T.J. Hollowood, Nucl. Phys. {\bf B352} (1991) 723}
\Rref{LS}{A.N. Leznov and M.V. Saveliev, Commun. Math. Phys.
{\bf74} (1980) 11; Lett. Math. Phys. {\bf3} (1979) 489\newline
P. Mansfield, Nucl. Phys. {\bf B208} (1982) 277; Nucl. Phys.
{\bf B222} (1983) 419\newline D. Olive and N. Turok, Nucl. Phys. {\bf
B257} [FS14] (1986) 277; Nucl. Phys. {\bf B265} [FS15] (1986)
469}
\Rref{BG}{A.B. Bilal and J.L. Gervais,  Nucl. Phys. {\bf B318} (1989)
579\newline O. Babelon, Phys. Lett. {\bf B215} (1988) 523}
\Rref{LSS}{D.A. Leites, M.V. Saveliev and V.V. Serganova, `{\sl
Embeddings of Osp($N/2$) and associated non-linear supersymmetric
equations\/}, Proc.~Third Yurmala Seminar (USSR 22-24 May 1985),
{\sl Group theoretical methods in physics\/} vol.~1 (VNU Science Press,
Utrecht, 1986) }
\Rref{BFK}{W. Boucher, D. Friedan and A. Kent, Phys. Lett. {\bf B172} (1986)
316\newline
P. Di Vecchia, J.L. Peterson and M. Yu, Phys. Lett. {\bf B172} (1986) 323}
\Rref{fend}{P. Fendley and K. Intriligator,  `{\sl Scattering and
thermodynamics in integrable $N{=}2$ theories\/}', Boston and Harvard
University preprint BUHEP-92-5, HUTP-91/A067, hep-th/9202011}
\Rref{sssm}{R. Shankar and E. Witten, Phys. Rev. {\bf D17}
(1978) 2148\newline
K. Schoutens, Nucl. Phys. {\bf B344} (1990) 665\newline
C. Ahn, D. Bernard and A. LeClair, Nucl. Phys. {\bf B346}
(1990) 409\newline P. Fendley and K. Intriligator, Nucl. Phys. {\bf
B372} (1992) 533}
\Rref{ntsm}{P. Fendley, S.D. Mathur, C. Vafa and N.P. Warner,
Phys. Lett. {\bf B243} (1990) 257}
\Rref{Serg}{V.V. Serganova, Math. USSR Iz. {\bf24} (1985) 539}
\Rref{KI}{T. Inami and H. Kanno, Commun. Math. Phys. {\bf 136} (1991)
519; Nucl. Phys. {\bf B359} (1991) 201}
\Rref{GPZ}{A. Gualzetti, S. Penati and D. Zanon, `{\sl Quantum
conserved currents in supersymmetric Toda theories}', University of
Milan preprint IFUM-426-FT}
\Rref{stod1}{H.C. Liao and P. Mansfield, Nucl.~Phys.~{\bf B344} (1990)
696; Phys.~Lett.~{\bf B267} (1991) 188;
Phys.~Lett.~{\bf B252} (1991) 230; Phys.~Lett.~{\bf B252} (1991) 237
}
\Rref{stod2}{H. Nohara and K. Mohri, Nucl.~Phys.~{\bf B349} (1991) 253;
\newline
S. Komata, K. Mohri and H. Nohara, Nucl.~Phys.~{\bf B359} (1991) 168
}
\Rref{stod3}{S. Penati and D. Zanon, `{\sl Supersymmetric,
integrable Toda field theories: the\/} $B(1,1)$ {\sl model\/}',
University of Milan preprint IFUM-421-FT
\newline
M.T. Grisaru, S. Penati and D. Zanon, Nucl.~Phys.~{\bf B369} (1992)
373; Phys.~Lett.~{\bf B253} (1991) 357
\newline
G.W. Delius, M.T. Grisaru, S. Penati and D. Zanon, Nucl.~Phys.~{\bf
B359}
(1991) 125; Phys.~Lett.~{\bf B256} (1991) 164
}
\endref
\ciao
